\documentclass[twocolumn,tighten]{aastex63} %
\usepackage{multirow}
\usepackage{amsmath}
\usepackage{graphicx}

\shorttitle{Rotation of the Solar Chromosphere}
\shortauthors{J. C. Xu, P. X. Gao \& X. J. Shi}

\begin{document}
\title{\textbf{On the Rotation of the Solar Chromosphere}} 
\correspondingauthor{Jing-Chen Xu}
\email{jcxu@ynao.ac.cn}

\author[0000-0002-8947-547X]{Jing-Chen Xu}
\affiliation{Yunnan Observatories, Chinese Academy of Sciences, Kunming 650011, China}
\affiliation{Center for Astronomical Mega-Science, Chinese Academy of Sciences, Beijing 100012, China}
\affiliation{Key Laboratory of Solar Activity, National Astronomical Observatories, Chinese Academy of Sciences, Beijing 100012, China}
\author[0000-0003-2921-1509]{Peng-Xin Gao}
\affiliation{Yunnan Observatories, Chinese Academy of Sciences, Kunming 650011, China}
\affiliation{Center for Astronomical Mega-Science, Chinese Academy of Sciences, Beijing 100012, China}
\affiliation{Key Laboratory of Solar Activity, National Astronomical Observatories, Chinese Academy of Sciences, Beijing 100012, China}
\author[0000-0002-5692-3395]{Xiang-Jun Shi}
\affiliation{Yunnan Observatories, Chinese Academy of Sciences, Kunming 650011, China}
\affiliation{Center for Astronomical Mega-Science, Chinese Academy of Sciences, Beijing 100012, China}
\affiliation{Key Laboratory of Solar Activity, National Astronomical Observatories, Chinese Academy of Sciences, Beijing 100012, China}


\begin{abstract}

Rotation is a significant characteristic of the Sun and other stars, and it plays an important role in understanding their dynamo actions and magnetic activities. In this study, the rotation of the solar chromospheric activity is investigated from a global point of view with an over 40 yr Mg II index. We determined the time-varying rotational period lengths (RPLs) with the synchrosqueezed wavelet transform which provides high temporal and frequency resolution; furthermore, we compared the RPLs with the photospheric and coronal RPLs obtained from the sunspot numbers and the 10.7 cm radio flux data. The significance of the RPLs is taken into consideration. We found that the RPLs of the chromosphere exhibit a downward trend, as do those that of the photosphere and corona; in addition, their RPLs at the recent four solar maxima also show a declining trend. It suggests that the rotation of the solar atmosphere has been accelerating during the recent four solar cycles, which is inferred to be caused by the declining strength of solar activity.
The variations of the solar atmospheric RPLs show periodicities of multiple harmonics of the solar cycle period, and it is modulated by the solar activity cycle.

\end{abstract}

\section{INTRODUCTION}
\label{sectIntro}

Rotation is one of the basic characteristics of the Sun as well as other stars. The interaction between rotation and convection leads to the dynamo action that is responsible for the solar activity cycle~\citep[e.g.,][]{Brun2017, Cameron2017}.
Due to the advancement of techniques and accumulation of data, there are increasing studies on the rotation of other stars~\citep[e.g.,][]{McQuillan2014, Benomar2018}.
A relationship named stellar gyrochronology is developed amongst the rotation rate (angular velocity of the solar surface, usually measured in ``degree/day" or ``$\mu$ radian/second"), color, and age in certain types of stars~\citep{Barnes2003, Barnes2007}.
Nonetheless, the Sun is the only star whose rotation can be investigated in detail, and the result might serve as a reference for the study of stellar rotations, especially that of solar-like stars.
It is well-know that the solar rotation is differential as a function of both latitude and depth.
It seems straightforward to determine the differential rotation rate of the Sun at first sight; however, there are still remarkable discrepancies after more than a century's work~\citep[e.g.,][]{Beck2000}.

Concerning the determination of the solar rotation rate, there are mainly three categories of methods used:  feature tracking (doppler or magnetic), doppler shifts, and helioseismology~\citep{Howard1984, Beck2000}.
The former two are meant for determining the rotation of the solar atmosphere (the photosphere and above), while the last one is used for probing the solar interior rotation beneath the photosphere~\citep{Howe2009}.
As the name implies,  the first method determines rotation rates by tracking magnetic features. A number of magnetic features in the solar atmosphere have been made using, for instance, sunspots, plages, faculae, coronal holes, coronal bright points, networks, giant cells, etc. Almost all of the features are believed to be linked with the local magnetic field. The second method is carried out by measuring the Doppler effect of spectral lines observed at the edge of the east and west solar disks at various latitudes.
Each of the two methods have shortcomings, and the results vary according to the methods and data that are chosen. For example, the tracer method is sensitive to the evolution of the specific features, while the spectroscopic method is susceptible to scattered light.

There is yet another way to estimate rotational periodicity, namely, the global-view method \citep{Mouradian2002, Heristchi2009, Li2011a, Li2012, Xie2017a}. This method makes use of various solar indices, and considers the rotation globally.
Determining the stellar periodicity with light curves is basically an application of the global-view method.
 While the tracer method is subject to the evolution and deformation of features,  the global-view method is not affected by such factors.
This method reflects the rate of the solar atmospheric variation independent of the quiet-Sun rotation. The fine structure, local layout, and short time evolution of individual structures are averaged out.

Based on the global-view perspective, the rotation rates of the solar photosphere and corona are studied extensively with various data and algorithms. The following are some of the recent results.
For example,
\cite{Heristchi2009} found a 52.4 yr period in the variation of the coronal rotation period lengths (RPLs; measured in ``days/360 degree" and often abbreviated as ``days") with the 10.7 cm radio flux (denoted with F10.7 afterwards), and that the global rotation rates of the sunspots and coronal flux are the same; they divided the data sets into two-year segments and estimated the RPLs with the maximum entropy method.
Based on the continuous wavelet transform (CWT), \cite{Li2011b} studied the RPLs of the sunspot numbers during 1849--2010; they found a linear secular decreasing trend by about 0.47 days, and there is no Schwabe cycle in the long-term variations of the RPLs.
Similarly, \cite{Li2011a} studied the sunspot areas during 1874--2010 with CWT, and also found a secular decreasing trend; besides, they found significant periods of 2.61 and 5.77 yr in the RPLs.
Later, \cite{Li2012} examined the yearly coronal RPLs obtained by \cite{Chandra2011} based on the daily adjusted F10.7 during 1947--2009; they found that the RPLs and the F10.7 each have a secular trend and there is a negative correlation between the two trends.
Also with F10.7 and the CWT method, \cite{Xie2017a} found that the coronal RPLs are varying with the solar cycle (SC) phase, and there is a phase difference of about 5.5 yr between the RPLs and F10.7 data sets.
\cite{Badalyan2017} explored the coronal RPLs with Fe XIV 530.3 nm flux during 1939--2011, and found that the variation of the equatorial RPLs of the corona is related with the even and odd cycle, which is not revealed in previous studies.
Recently, \cite{Deng2020} studied the coronal RPLs with the modified coronal index during 1939--2019 and the CWT method; they found a decreasing trend during the time interval  1939--2019, which confirms the results of coronal rotation by \cite{Li2011a, Li2011b, Xie2017a}.
The rotation of the photosphere is well determined, while that of the corona remains more controversial, which is probably due to the  density and temperature structure of  the corona.

The previous investigation of the rotation of the Sun is mostly concentrated on the solar photosphere and corona; however, the rotation of the chromosphere, a layer lying just above the photosphere (at which the temperature is believed to increase outwards) and below the corona, which is probably critical for the corona heating problem, is rarely studied and less understood.
In the solar chromosphere, the magnetic field dominates the structuring of the low-$\beta$ plasma, and the temperature is relatively stable except at the very top, thus the variation of density and temperature is not that dramatic compared with the corona~\citep{Priest2014a}.
Since the chromosphere is likely rotating more rigidly, studying the rotation of the chromosphere with the global-view method should be more reliable.
Decades ago, \cite{Antonucci1979, Antonucci1979a} studied the relation between lifetime of feature tracers and their sizes in chromospheric rotation; they found that short-lived features of Ca II K  rotate at the same rate as the chromospheric plasma, i.e. faster than the photospheric plasma, while long-lived features rotate like the green corona or coronal holes, almost rigidly.

In previous studies using the global method,  the auto-correlogram or the wavelet technique are often used. There are several setbacks with these applications that may cripple the results.
To begin with, in works that use the CWT, the rotation period of each time is often simply determined by selecting the point with the largest power at that time. In this case, the rotation rates may vary abruptly from day to day, which is obviously not true since the rotation rate should vary continuously. \cite{Hasler2002} have showed that magnetic features at high latitudes may appear and disturb the determination of the RPL; often several distinct periods appear almost simultaneously. Furthermore, in the auto-correlation method, in order for the method to be valid, the time series has to be divided into segments with an artificially chosen length. In that case, the obtained RPL is actually an average value during a long period of time. Besides, the time resolution is low. Last but not least, the significance of the RPLs is rarely taken into consideration, which may lead to erroneous results in the subsequent analyses.

The MgII index is a measure of the chromospheric activity that is fairly easy to measure and relatively insensitive to instrumental artifacts. It is a dimensionless quantity that has an advantage of being independent of the absolute calibration and aging of the instruments~\citep{Heath1986}.
The emission core of the Mg II is formed at a temperature of  $\sim$7000 K in the upper chromosphere, while the broad absorption feature of the Mg II wing comes from the upper photosphere; the latter is much less variable. The variability of the photosphere is quite small, while the chromospheric emission varies by about 30\%  as active regions evolve and come in and out of view of the solar disk~\citep{Snow2005}.
In comparison with the Mg II index, the sunspot numbers is a coarse measurement of solar magnetic variability. The Mg II index is also one of the longest records of solar variability reaching back more than 40 yr.

The above reasons justify a further investigation of the topic and our present analysis.
In this paper, we will explore the variation of the RPLs of the solar chromosphere with the Mg II h k core-to-wing ratio, i. e., the Mg II index, based on the global-view method.
In Sect.~\ref{sectData}, we will introduce the Mg II index and the data. Sect.~\ref{sectMtd} includes the methods and Sec.~\ref{sectAnR} the analyses and results. Sect.~\ref{sectDiscu} concludes the paper and discusses the results.


\section{Data}
\label{sectData}

\begin{figure}
 	\centerline{\includegraphics[width=0.95\linewidth, clip=]{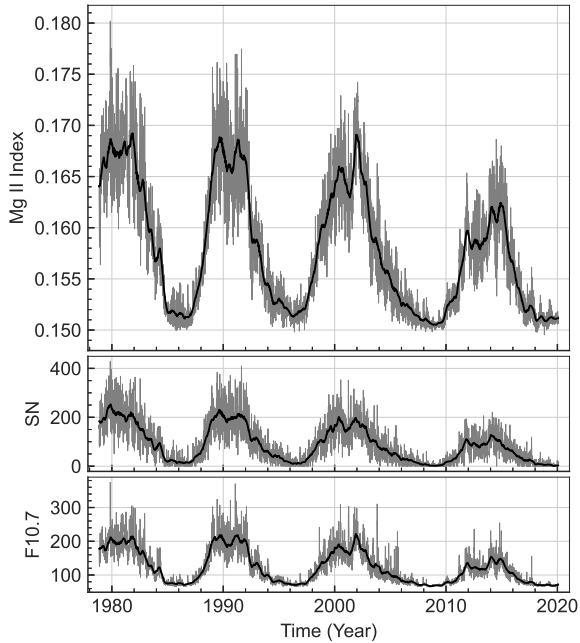}}
	\caption{Data sets of the Mg II index (upper panel), the sunspot numbers (SN; middle panel), and the 10.7 cm radio flux (F10.7; bottom panel). In each panel, the gray curve shows the daily values, and the thick black curve indicates the 91-day moving averaged data. All three of the data sets span from 1978 Nov 7 to 2020 Jan 30. \label{fig-data}}
\end{figure}

The Mg II core-to-wing ratio was proposed and developed by~\citet{Heath1986}.
It is an important measure of the variability of radiation coming from the solar chromosphere. As stated previously, it does not tend to be subject to the precision of absolute measurement or instrumental degradation.
For a low-resolution (e.g., 1.1 nm) instrument, the definition of the Mg II core-to-wing ratio can be written as:
\begin{equation}
I = \dfrac{4[E_{279.8} + E_{280.0} + E_{280.2}]}{3[E_{276.6} + E_{276.8} + E_{283.2} + E_{283.4}]}
\end{equation}
where $E$ means the emission by the wavelength indicated in the subscript, and $I$ is the ratio.
There are other algorithms, but in any cases the MgII index is a ratio of the chromospheric contribution to the photospheric contribution.
The Mg II Index is constructed by combining various Mg II core-to-wing ratio data sets.
It is a suitable proxy for solar activity; particularly, it has a good correlation with the EUV emission which is important for understanding the upper solar atmosphere~\citep{Snow2005}.
It is a more commonly available measurement which can be used to represent the true EUV irradiance \citep{SnowMartin2014}, and it has been widely used in EUV, UV, and total solar irradiance models~\citep{Viereck2004}.

In this study, we use the Bremen composite Mg II index time series, which is shown in the upper panel of Figure~\ref{fig-data}. The data sets of sunspot numbers (SN) and coronal radio flux at 10.7 cm wavelength (F10.7) are also displayed in the middle and bottom panels. The time interval of all of the three series is from November 7, 1978 to January 30, 2020 (15061 days in all). The gray curves show the daily values, and the superimposed black curves show the corresponding time series smoothed with a 91-day moving average filter.
The Mg II index clearly shows a $\sim$11-yr period variation along with the solar activity cycle.
The solar minimum to maximum amplitude of the Mg II index is 12.3\% of the average minimum as derived from the 91-day filtered data.

\begin{figure}
 	\centerline{\includegraphics[width=0.95\linewidth, clip=]{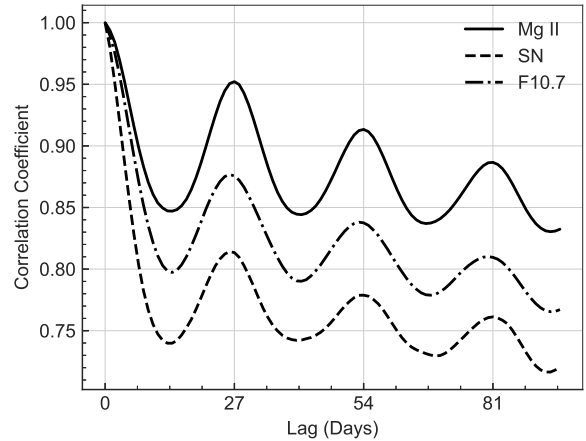}}
	\caption{Auto-correlogram of the Mg II index (solid line), SN (dashed line), and F10.7 (dashed-dotted line) data sets.   \label{fig-acorr}}
\end{figure}

Figure~\ref{fig-acorr} is a plot of the auto-correlogram of the Mg II index at a time scale of several rotation period lengths (RPLs), as well as that of SN and F10.7.
Not surprisingly, the wax and wanes of the three curves obviously depict an $\sim$27-day periodicity, which is caused by the rotation of the Sun. Sunspots come in and out of the visible solar disk as the Sun rotates. If an auto-correlogram with large lags is performed, the Schwabe cycle can be revealed as well.
It is obvious in the figure that the correlation coefficients of the Mg II index in maxima at around lags of multiples of 27 days are larger than that of the two others'. That is to say, the Mg II index shows a more notable rotational characteristic. In the next section, we will study the RPLs of the Mg II index, as well as its temporal variation.

\section{Methods}
\label{sectMtd}

In this work, we use the Synchrosqueezed wavelet transform (SWT) to determine the RPLs rather than the CWT.
 SWT is a time-frequency analysis method based on the well-known wavelet transform. It is useful for analyzing signals that is composed of multiple oscillatory modes. We will extract the rotational signal in the Mg II index, SN, and F10.7 with SWT, and study their temporal variation.

In terms of signal extraction, it is inevitable to mention the EMD method proposed by~\citet{Huang1998}.
The EMD method is a data-driven adaptive sifting algorithm that aims to decompose a signal into a reasonably small number of components that represent the intrinsic high- and low-frequency oscillations in the original signal.
A complete description of this algorithm can be found in~\citet{Huang1998}.
The EMD has shown its broad usefulness in a wide range of applications including astrophysics, for instance, \citet{Li2012, Lee2015, Lee2016, Deng2015, Kolotkov2016, Xiang2016, Gao2016, Gao2017a, Xu2018}.
In order to resolve the ``mode mixing" problem in the original EMD method (and new problems brought out), two derivative methods, i.e. the Ensemble Empirical Mode Decomposition (EEMD) and the Complete Ensemble Empirical Mode Decomposition with Adaptive Noise (CEEMDAN), are proposed by~\citet{Wu2009} and ~\citet{M.E.Torres2011}, respectively.
Nevertheless, the mode mixing problem still cannot be completely solved with these improvements.
Furthermore, EMD is essentially a numeric and empirical algorithm that poses challenges to a mathematical understanding of the approach as well as the results produced.

Recently, \cite{Daubechies2011} developed the SWT method as an alternative to the EMD scheme for separating a nonstationary signal with time-varying amplitudes and instantaneous frequencies into a superposition of well-defined frequency components.
SWT is a combination of wavelet analysis and reallocation method which captures the flavor and philosophy of the EMD approach, albeit using a different approach in constructing the components.
It introduces a precise mathematical definition for a class of functions that can be viewed as a superposition of a reasonably small number of approximately harmonic components.
The process of using SWT to decompose a signal $f(t)$ into a series of intrinsic mode type components can be described by the following four steps.

\begin{enumerate}
\item Get the continuous wavelet transform $W^{\psi}_f$ of the signal $f(t)$:
$$W^{\psi}_f(a, \tau)=\int_{-\infty}^{\infty}f(t)a^{-1/2}\psi^{*}(\dfrac{t-\tau}{a})dt.$$
Here, $\psi$ is an appropriate mother wavelet (e.g., the morlet wavelet),  $a$ is the scale stretch, and $\tau$ is the time shift. $\psi^{*}$ denotes the complex conjugate of $\psi$.

\item Then, take the phase transform $\omega_f(a, \tau)$, defined as the derivative of the complex phase of $W^{\psi}_f(a, \tau)$. Extract instantaneous frequency from the output of CWT:
$$\omega_f(a, \tau)= \dfrac{1}{2\pi i} \dfrac{\partial_\tau W^{\psi}_f(a, \tau)}{W^{\psi}_f(a, \tau)}.$$
$\partial_\tau$ means the partial derivative of  $\tau$.

\item Perform the synchrosqueezed transform:
$$T_f(a, \tau) = 2\Re{\dfrac{1}{C'_\psi}\int_{b}W^{\psi}_f(b, \tau)\delta(\tilde{\omega}_f(a,\tau)-b)\dfrac{db}{b}},$$
where $\Re$ denotes the real part of a complex number and $C'_\psi = \int^{\infty}_0 \hat{\psi}^*(\xi)\frac{d\xi}{\xi}$.
In this step, the influence of $\psi$ from the CWT is removed and the localized frequency information that we want is ``encoded''. The coefficients of the CWT in the time-scale plane are reassigned according to the ``map'': $(a, \tau)$ to $(\omega_f(a, \tau), \tau)$.

\item The reconstruction of $f_k$ is through:
$$f_k(t)=\int_{|1-a\phi^{'}_k(t)|<d}T_f(a, \tau)da.$$
Since the (complex) coefficients are moved only along the scale axis, the transform remains invertible.
\end{enumerate}

This method has been used in, for instance,  investigating the midscale period of solar mean magnetic field~\citep{Feng2017} and the periodicity of polar faculae~\citep{Deng2020}.
We will use SWT to extract the rotational signal of Mg II index, SN, and F10.7.

\begin{figure*}
 	\centerline{\includegraphics[width=0.95\linewidth, clip=]{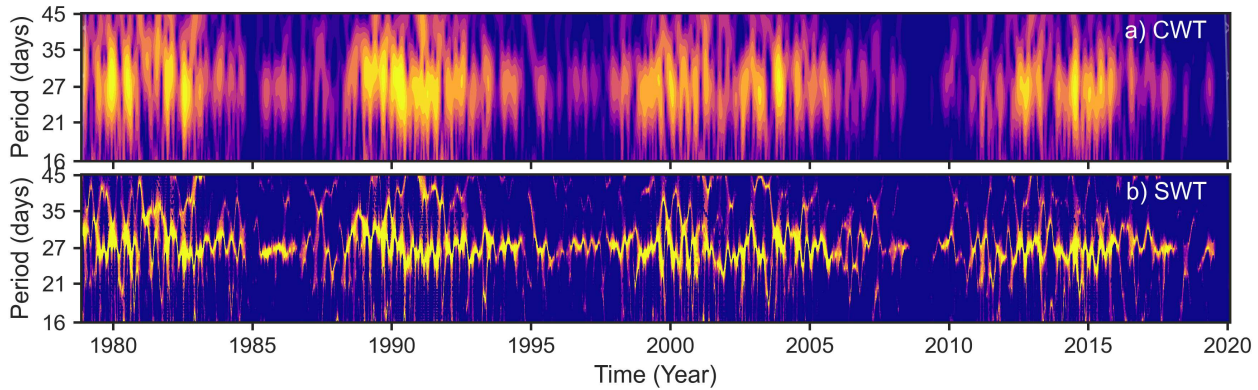}}
	\caption{Wavelet power spectra of the Mg II index derived from (a) CWT and (b) SWT, respectively. Both of them are based on the morlet wavelet. The power at periods between 16 and 45 days are displayed. Compared with the CWT spectrum, the SWT spectrum is much more concentrated in the ridges. The rotational period lengths are extracted along the ridge in the SWT spectrum.
\label{fig-swt_cwt}}
\end{figure*}

\section{Analyses and Results} 
\label{sectAnR}

In order to illustrate the advantage of the SWT method, the power spectra of the Mg II index based on both CWT and SWT, respectively, are shown together in Figure~\ref{fig-swt_cwt}. The morlet wavelet is used in both cases. To clearly show the rotational signal, only spectra at the period range between 16 and 45 days are displayed. Since the wavelet technique can be well localized in both time and frequency, there is no need to modify the original data set which may cause false signals, such as detrend (subtracting a smoothed curve) or normalize (subtracting the mean value and dividing by the variance).
As can be seen on the time-period (or time-frequency) plane, the SWT spectrum is obviously sharper and more concentrated around frequency ridges; in other words, it has a much higher frequency resolution than that of CWT.  In the SWT spectrum, the temporal variation of the rotational signal, which reflects the chromospheric rotation rates is roughly shown.

\begin{figure}
   \centerline{\includegraphics[width=0.95\linewidth, clip=]{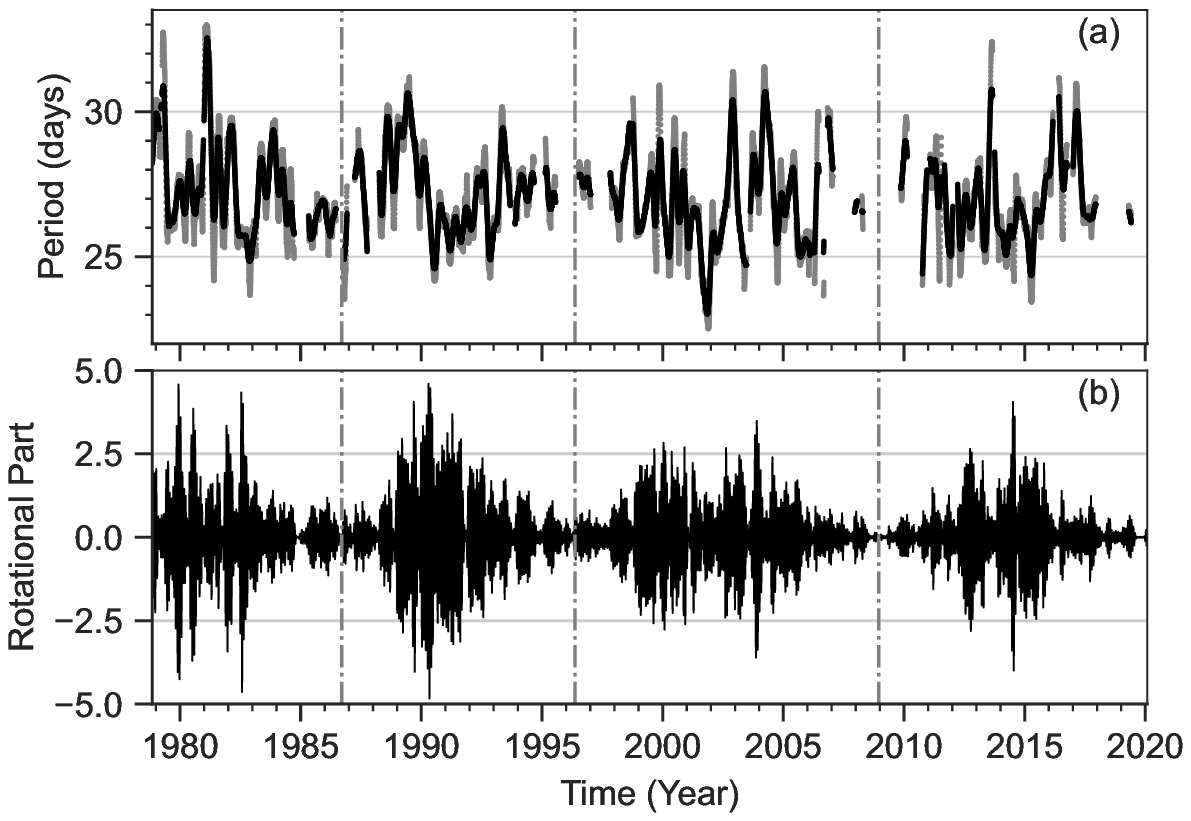}} %
   \centerline{\includegraphics[width=0.95\linewidth, clip=]{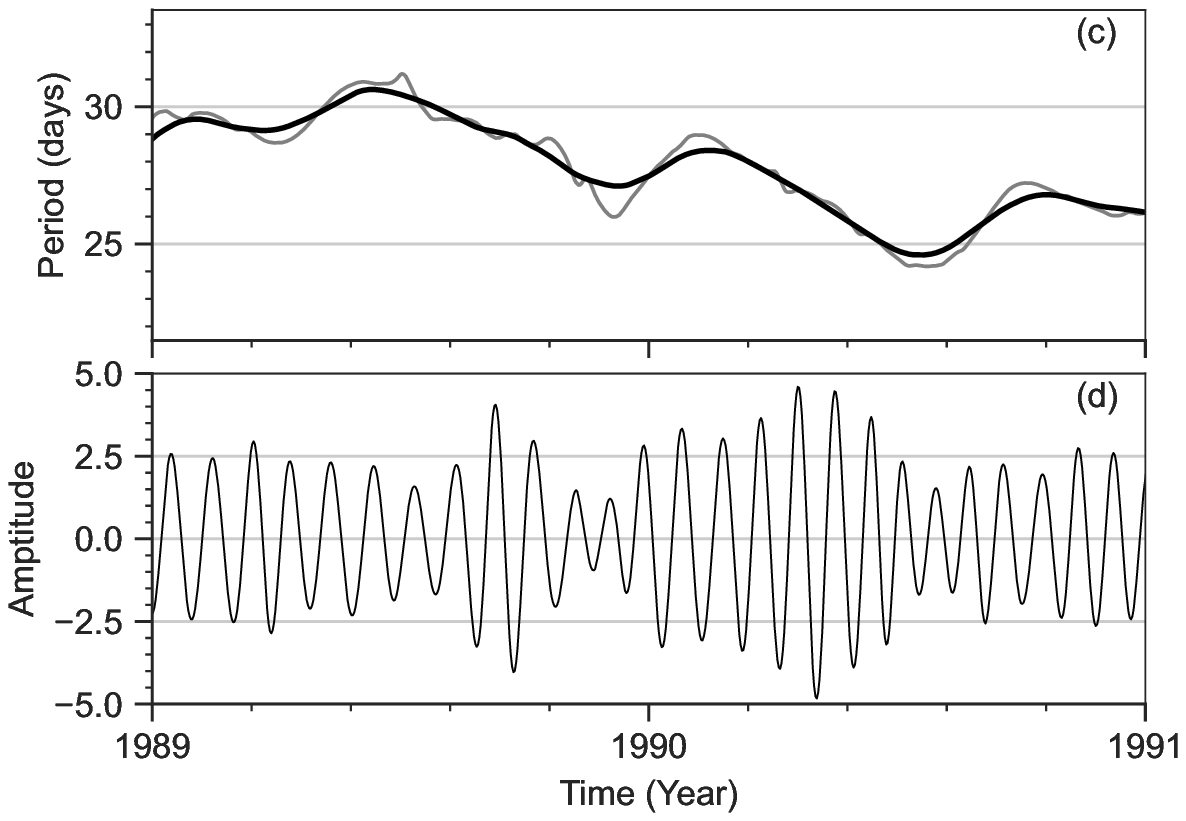}} %
 \caption{The rotational period ridge and the oscillatory component of the rotational mode in the Mg II index obtained from SWT. (a) The daily significant rotational period lengths (gray dotted line), as well as the 390 day smoothed line (black dotted line); they are both discontinuous due to the removal of insignificant periods. The vertical dashed-dotted lines indicate dates of solar minima. (b) The reconstructed rotational signal component.
 (c) and (d) Zoomed in images of (a) and (b), respectively,  between the years of Jan 1, 1989 and Jan 1, 1991 (during solar maximum of SC 22), which is meant to show the RPLs and rotational signal clearly. During that time, all of the RPLs are significant and thus the curves are continuous.
  \label{fig-ridge}}
\end{figure}

To further study the RPL variation quantitatively, we can obtain the time-variable RPLs by extracting the approximately continuous ridge at around $\sim$27-day period.
 In previous studies that based on the wavelet technique, the RPLs are obtained by just  selecting the frequency corresponding to the largest power at each time point (in this case, each day). It is possible that the RPL may vary abruptly from day to day, which is unrealistic, assuming that the length of the rotation rate of the solar atmosphere changes continuously.
In all, because the RPLs are extracted along the ridge, the result will not be jeopardized by abrupt jumps. Besides, the range used here is 22--33 day, which is a wider range than 25--31 as in \citet{Li2011a} and \citet{Xie2017a}. As shown later in Figure~\ref{fig-ridge} (e), only a small part of the RPLs reach the edge of the range. Note that all of the RPLs are synodic periods here.

\begin{figure}
   \centerline{\includegraphics[width=0.95\linewidth, clip=]{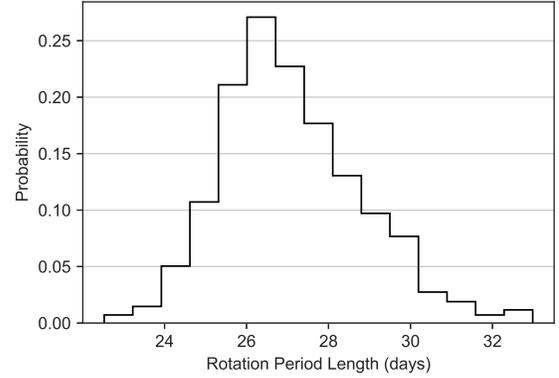}} %
 \caption{The distribution of the significant rotational period lengths (RPLs) of the Mg II index, i.e., the distribution of the data is shown in Figure~\ref{fig-ridge} (a).
  \label{fig-distri}}
\end{figure}

\subsection{Rotational period length}

\begin{figure*}
 	\centerline{\includegraphics[width=0.95\linewidth, clip=]{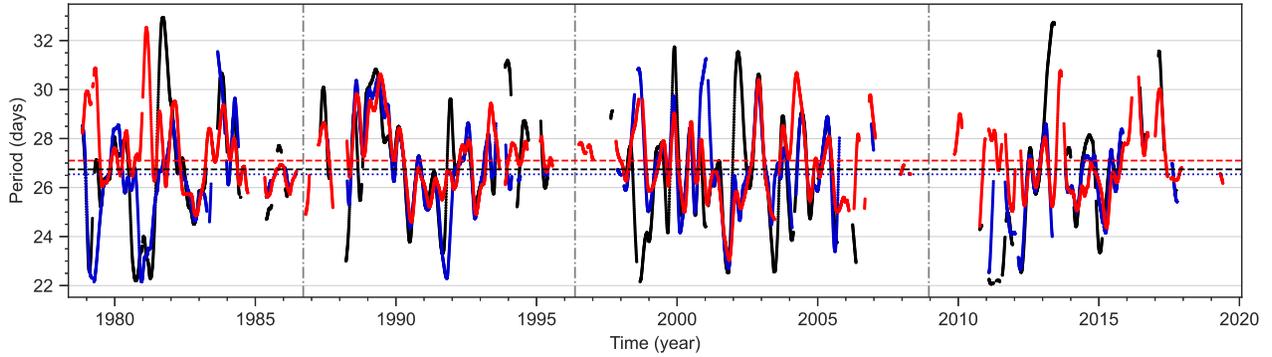}} %
	\caption{The time-variable significant rotational period lengths (RPLs) of the Mg II index (red dots), sunspot numbers (black dots), and F10.7 flux (blue dots). The vertical dashed-dotted lines indicate dates of solar minima. The three horizontal colored dashed lines indicate the corresponding mean values of the three RPLs. \label{fig-3ridge}}
\end{figure*}

\begin{figure}
   \centerline{\includegraphics[width=0.95\linewidth, clip=]{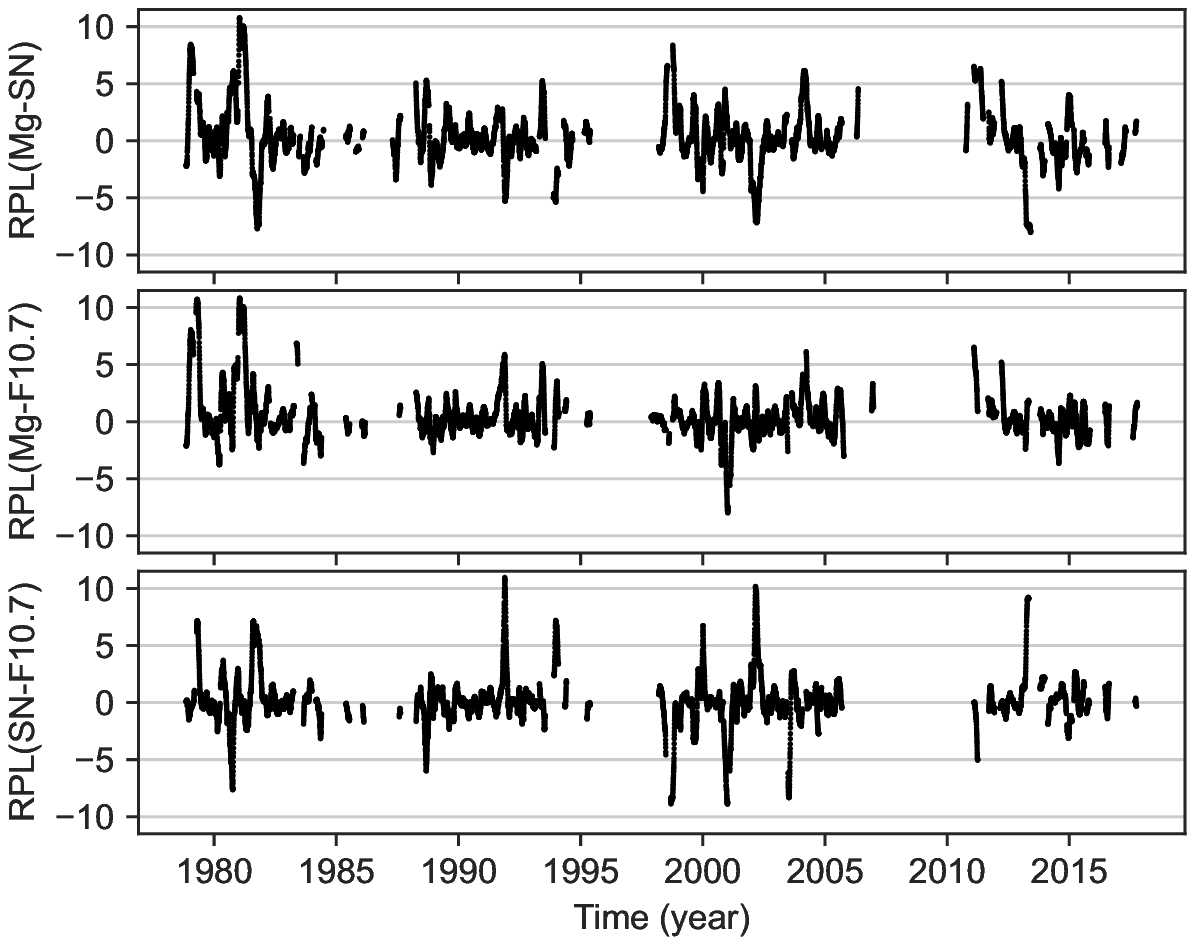}} %
 \caption{Differences between each two of the three RPLs (indicated in the y-axis label). Only significant RPLs are taken into consideration in calculating the differences. There are no systematic patterns in the differences. \label{fig-diff_RPL}}
\end{figure}

Based on the SWT spectrum of the Mg II index in Figure~\ref{fig-swt_cwt}, we obtained the ridge of the $\sim$27-day rotational signal. Besides, the oscillatory mode along the $\sim$27-day ridge is also extracted.
Figure~\ref{fig-swt_cwt} shows that the wavelet power is at a very low level sometimes, especially at solar minima, suggesting that the periodic signal is not distinguishable when the Sun is quiet. Therefore, unlike previous studies based on the wavelet technique, we take the significance of the RPLs into consideration.
The steps are as follows:
\begin{itemize}
  \item determine the value of RPLs along the $\sim$27-day ridge and extract the oscillatory signal with SWT;
  \item get the areas above the 95\% significance level as a function of time and period with CWT;
  \item check whether each of the RPLs lies within the corresponding significance area according to its location on the time-period plane;
  \item abandon all RPLs that are ``not significant''.
\end{itemize}
The significant RPLs of the Mg II index and its $\sim$27-day rotational component are shown in Figure~\ref{fig-ridge}.
It is obvious in Figure~\ref{fig-ridge} (b) that the amplitude of the rotational signal is much stronger at solar maxima than that at solar minima. During the times when the rotational signal is very weak, the RPLs are unable to be determined.
The activities of the chromosphere is generally linked with magnetic fields residing  downwards, i.e., in the photosphere and below. When there are active regions passing by the visible disk, the rotational signal tends to be more prominent.
Figure~\ref{fig-ridge} (a) indicates that the insignificant periods are mainly distributed in solar minima during the times when there are less, or even no, active regions.
That does not mean that the rotation of the Sun is unable to be determined during solar minima---it just cannot be determined from the signal in the Mg II index when the signal is too weak, not to say SN and F10.7 flux which show less prominent rotational signals.
Analyses and inferences based on the insignificant RPLs may be erroneous.
The percentages of insignificant RPLs are substantial; the numbers of RPLs above the 95\% significance level amount to 10829 (71.9\%), 8409 (55.8\%), and 8297 (55.1\%) for the Mg II index, SN, and F10.7, respectively. Again, the Mg II index proves itself a better proxy for estimating solar rotation.

To show the temporal variation of periods and rotational signal, a zoom in of  Figure~\ref{fig-ridge} (a) and (b) in the years 1989 and 1990 (during the solar maximum of SC22) is displayed  in Figure~\ref{fig-ridge} (c) and (d) to give a clear view; during that time, all of the RPLs are significant and thus the curves are continuous.
According to the extracted ridge, the rotational period of the Mg II index is  $27.10\pm1.72$ days within the time interval considered. The distribution of Mg II index RPLs is shown in Figure~\ref{fig-distri} (the distributions of the RPLs of SN and F10.7 are similar). Most of the RPLs are distributed around 27 days,  and only a few RPLs reach the edge of the range (22 or 33 days).

With the same procedure, we extracted the RPLs of the SN and F10.7, and Figure~\ref{fig-3ridge} shows a comparison of the variations of the three RPLs.
It can be seen that the three RPLs match only occasionally.
To give a detailed comparison, the daily differences between each two of the three RPLs are shown in figure~\ref{fig-diff_RPL}.
Specifically, the mean and standard deviation of the differences of RPLs between Mg and SN is 0.24$\pm$2.53 days, that between Mg and F10.7 0.47$\pm$2.07 days, and that between SN and F10.7  0.16$\pm$2.21 days. The standard deviations are too large to claim definitively which one is larger, and the differences exhibit no systematic patterns.
It reveals that the three RPLs are different during most of the time, which means that the rotation rates of the three solar atmospheric layers are different from each other during most of the time.
Though activities in the three atmospheric layers are believed to be coupled via magnetic fields and thus their RPLs should agree with each other, their rotation rates actually show substantial discrepancies.

\begin{figure}
   \centerline{\includegraphics[width=0.95\linewidth, clip=]{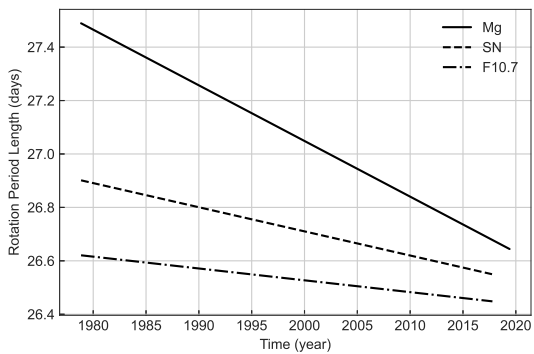}} %
 \caption{Linear trends of the three RPLs. The original data points are not plotted so as to show the slopes more clearly. \label{fig-linear_trend}}
\end{figure}

\subsection{Linear trends}

The long-term variation of the rotation rate of the Sun has always been an interesting issue.
After we obtained the three time-varying RPLs, we fitted each of them with a linear function. In the fitting, only the significant periods were taken into consideration. The results are displayed in Figure~\ref{fig-linear_trend}.
It is obvious that all three of the RPLs show a downward trend. The slope of the Mg II RPLs trend is the most steep one among the three. The lengths of the rotation periods are decreasing, that is to say, the rotation of the Sun is accelerating overall during the recent forty years.

This decreasing trend agrees with the results of previous studies on the long-term variation of the photospheric and coronal rotation rates. For example,
\cite{Li2012} investigated the long-term variation of the coronal rotation based on the RPLs result obtained with auto-correlation by \cite{Chandra2011} and found a weak decreasing trend during 1947--2009; \cite{Xie2017a} confirmed this result with the CWT.
Recently, \cite{Deng2020} studied the coronal rotation rate  with the modified coronal index during 1939--2019, and also found that the coronal RPLs exhibit an obvious decreasing trend.

In all, despite the fact that the three RPLs generally differ from each other, all of them show a decreasing trend. We speculate that this is due to the weakening of the global solar magnetism.
During the recent four SCs, the amplitude of the SN is decreasing. There are less strong magnetic flux tubes emerging in the solar atmosphere.
It has long been found that big sunspot groups rotate slower than small sunspots~\cite{Howard1984}.
Strong magnetic fields tend to repress the surface rotation~\citep{Xu2016}.
When there are less strong magnetic fields emerge out from under the photosphere, the rotation of the whole solar atmosphere accelerates.
Note that if the significance of the RPLs is not taken into consideration, the trend of the SN RPLs is different---it shows an upward trend instead.

 \begin{figure}
  	\centerline{\includegraphics[width=0.95\linewidth, clip=]{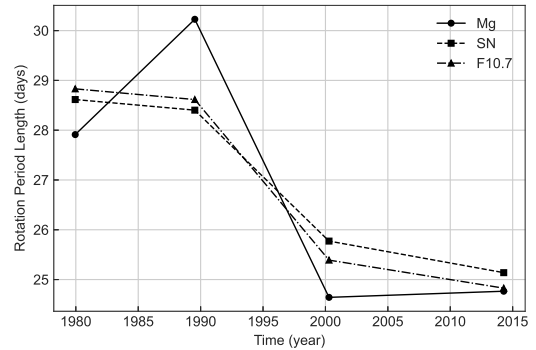}} 
 	\caption{Variations of the three RPLs at the recent four solar maxima. \label{fig-max_var}}
 \end{figure}

Figure~\ref{fig-max_var} shows only the RPLs at the recent four solar maxima.
From SC21 to SC24, the variation of both SN and F10.7 RPLs at maxima show a similar monotonous linear trend. The behavior of the Mg RPLs at maxima is different, but it also shows an overall decreasing trend.
Some studies (e.g., \citealt{Brajsa2006,Xie2017a}) found that the RPLs are relatively larger during solar minima. However, during solar minima, the rotation signals are mostly insignificant and the RPLs cannot be determined, thus we cannot compare the values of RPLs between solar minima and maxima here.

\subsection{Periodicity of the RPLs}

\begin{figure}
   \centerline{\includegraphics[width=0.95\linewidth, clip=]{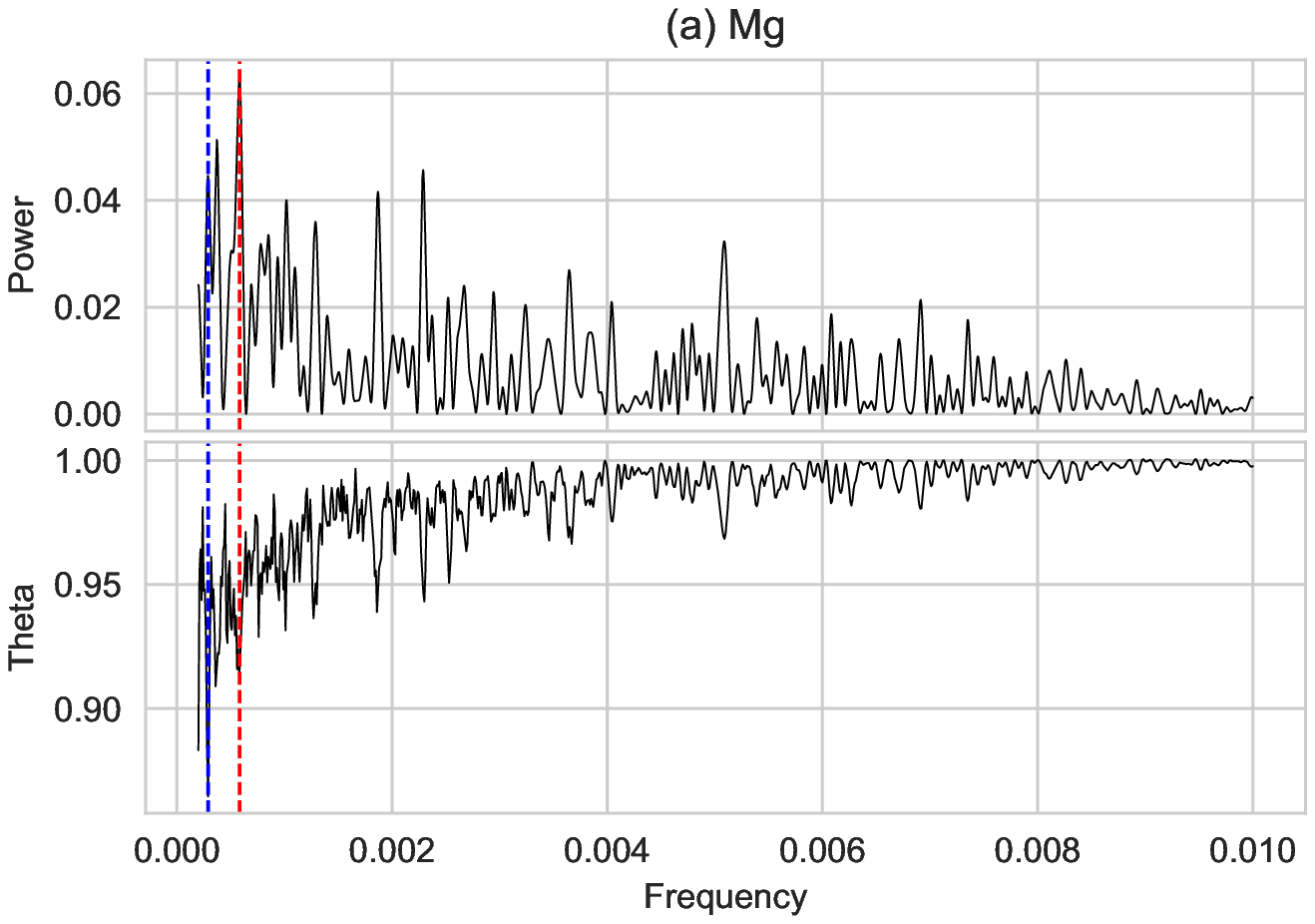}}
 \centerline{\includegraphics[width=0.95\linewidth, clip=]{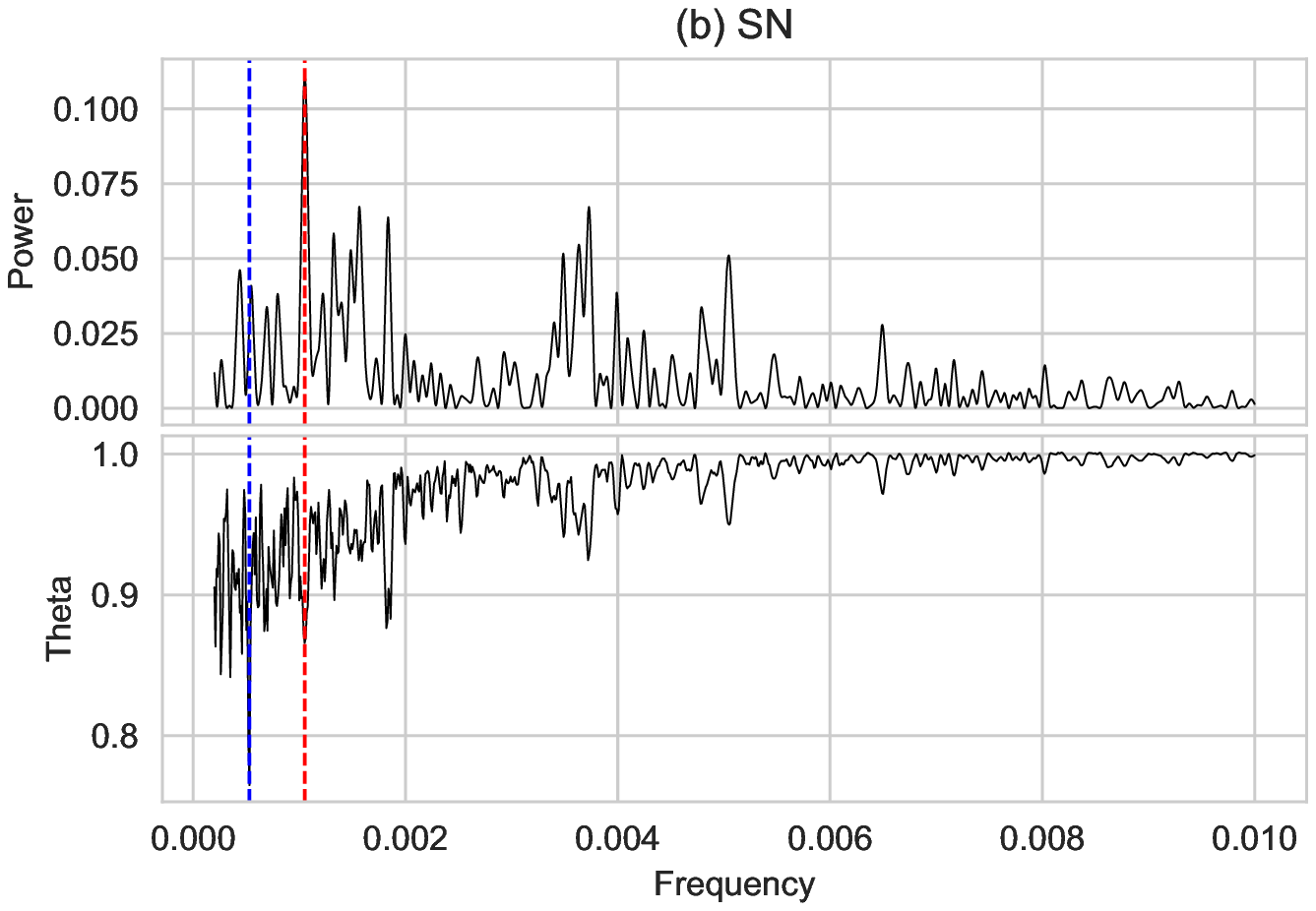}}
 \centerline{\includegraphics[width=0.95\linewidth, clip=]{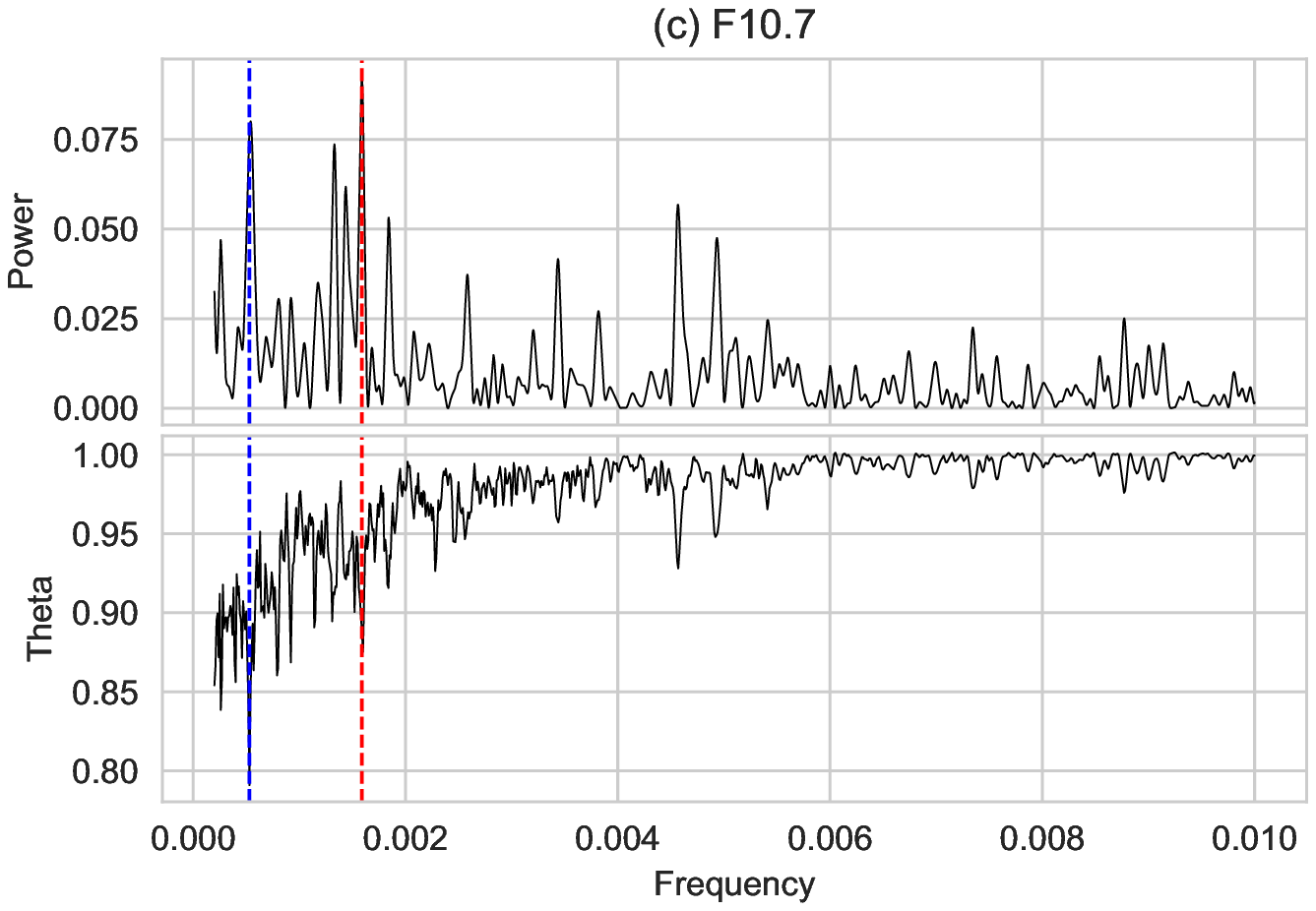}} %
 \caption{LSP and PDM (see the text) results of the three RPLs: (a) RPLs Mg, (b) RPLs SN, and (c) RPLs F10.7. In each subfigure, the red dashed  vertical line indicates the frequency corresponding to the maximum LSP power, and the blue dashed vertical line indicates the frequency corresponding to the minimum theta of PDM.  \label{fig-lsp_pdm}}
\end{figure}

Now that we have the variations of RPLs, it is natural to check whether they are periodic.
After excluding the insignificant periods that mainly occur during solar minima, the RPL data sets are not continuous. They can be considered unevenly spaced time series. Therefore, we use the Lomb-Scargle periodogram (LSP) and the phase dispersion minimization (PDM) method to study their periodicities. While the former method is essentially fitting the signal with sinusoidal functions, the later is also suitable for potentially nonsinusoidal signals~\citep{Zechmeister2009, Xu2016}. We use both methods for the analysis here so as to avoid false periods.
The results of LSP and PDM are shown in Figure~\ref{fig-lsp_pdm} and Table~\ref{tbl-1}. For each of the three RPLs, the results are shown together in the upper (LSP) and lower (PDM) panels of each subfigure, respectively.

\begin{table}[htp]
\caption{The frequencies (periods) of the three RPLs; they are calculated with both LSP and PDM as shown in Figure~\ref{fig-lsp_pdm}.}
\begin{center}\begin{tabular}{c  c c c c}
\hline
RPLs & \multicolumn{2}{c}{LSP} & \multicolumn{2}{c}{PDM} \\
\hline
 & Frequency & Period & Frequency & Period \\
 & (1/day)  &  (yr) & (1/day)  &  (yr)\\
 \hline
Mg  & $5.82\times10^{-4}$  & 4.71  & $2.90\times10^{-4}$  & 9.45  \\
SN & $1.05\times10^{-3}$  & 2.61  & $5.30\times10^{-4}$  & 5.17  \\
 F10.7 & $1.59\times10^{-3}$  & 1.73  & $5.30\times10^{-4}$  & 5.17  \\
\hline
\end{tabular}\end{center}\label{tbl-1}\end{table}%

The peaks of LSP power generally agree well with the minima of PDM theta which is the statistical variable of PDM~\citep{Stellingwerf1978}. Though for each one of the RPLs, the corresponding frequency of the highest LSP power does not match the corresponding frequency of the lowest PDM theta, harmonic relations are inferred between them. The PDM periods are inferred to be the 2 (for Mg II and SN) or 3 (for F10.7) multiple harmonics of the LSP periods.
In the calculation, the period resolution of LSP is $\sim$0.01 yr, and that of PDM is $\sim$0.05 yr.

The RPLs of the Mg II index shows a period of 9.45 yr, we infer that it is corresponding to the Schwabe cycle. Both SN and F10.7 show a period of 5.17 yr; we infer it is the 1/2 harmonic of the Schwabe cycle.
The  2.61 yr period of SN RPLs variation agrees well with \citet{Li2011a} who also found a significant period of 2.61 yr in the variation of photospheric RPLs from sunspot areas during 1874--2010.
\cite{Xie2017a} found a significant period of 10.3 yr in coronal rotation rates based on CWT and the F10.7 data during 1947--2014; the 5.17 period found here is likely the 1/2 harmonic of it. The shorter period lengths are probably caused by the shorter data length.
In contrast, \cite{Li2012} found that there is no $\sim$11-yr Schwabe cycle of statistical significance for the secular coronal rotation variation.
The periodicity of the rotation rate variation may be spoiled by the method and data that are chosen, which leads to the discrepancies between our result and previous results.

\subsection{Solar cycle modulation on the RPLs}

 \begin{figure}
  	\centerline{\includegraphics[width=0.95\linewidth, clip=]{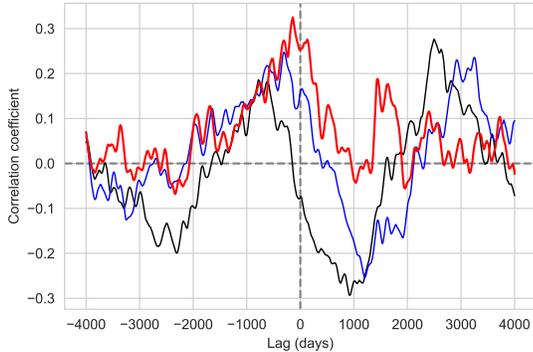}} %
 	\caption{Cross-correlogram between RPLs obtained from Mg (red), SN (black), and F10.7 (blue), respectively, with the data set of SN. The abscissa indicates the time shift of the RPLs against SN, and positive values representing forward shift.  \label{fig-xc_SN}}
 \end{figure}

To investigate whether the variation of RPLs is related with the solar activity cycle, we plotted the cross-correlogram between each of the three RPLs and the daily SN dataset in Figure~\ref{fig-xc_SN}. In the figure, the abscissa indicates the lag of the RPLs with respect to SN, with positive values representing forward shift.
In the calculation, only RPLs above the 95\% significance level are used, and the SN data set is filtered with a 91 day window. The specific method for calculating correlation coefficients between data sets with data gaps is described in~\citet{Xu2016}.

The wax and wane of the correlation coefficients indicates that all of the three RPLs are connected with the solar activity represented by SN.
All of the three RPLs lag behind the SN in time phase. Specifically, RPLs-Mg and RPLs-F10.7 lag SN by 140 and 306 days (5.2 and 11.3 rotation periods), respectively, and the RPLs-SN lag by 781 days (2.1 yr) in time phase.
The result between RPLs-SN and solar activity is similar to that found in \citet{Xie2017a}, though there are differences in the exact values of lag.
From the periodicities of the RPLs and their phase relation with the SN, we infer that the RPLs are modulated by the solar activity cycle.

\section{Conclusion and Discussion} \label{sectDiscu}

We investigated the rotation of the chromosphere and its variation during the recent 40 yr (1978--2020) in a global view, and compared the result with that of the photosphere and corona. The global-view method provides a feasible and exclusive way to track the continuous variation of solar rotation rates with a high temporal resolution. The rotational period lengths (RPLs) are extracted from the Mg II index, the sunspot numbers (SN), and the 10.7 cm radio flux (F10.7) data sets via the SWT method, which has a much higher frequency/period resolution than the CWT. Besides, we take the significance of the RPLs into consideration, which improved the reliability of the results.
We studied the long-term trends of the RPLs, their periodicities, and relationships with the solar activity.

The RPLs of the three solar atmospheric layers and their differences reveal that their global rotation periods only agree occasionally. Their rotation rates are different during most of the time, and there is no systematic pattern in the differences between any two of them.
The average values of the differences are in the range of 0.16--0.47 days, and the standard deviation is 2.07--2.53 days. We admit that the results may be affected by the absence of a significant portion of RPL values during solar minima.
\citet{Heristchi2009} studied the global rotation rates of the photosphere and corona with SN (from SC9 to SC23) and F10.7 (from SC19 to SC25) data. They divided the data sets into 2 yr segments and analyzed them with the maximum entropy method; then, they found that the solar rotation unfolding during the cycle is complex and the photosphere and corona show the same global rotation rates.
The rotational rate variation of the three atmospheric layers during the cycle revealed here is indeed complicated, which agrees with \citet{Heristchi2009}.
Besides, the variation of rotation rates of the photosphere and corona at solar maxima do agree well, as shown in Figure~\ref{fig-max_var}. However, the global rotation rates of  the photosphere and corona are seldom the same. As indicated in Figure~\ref{fig-3ridge}, it is obvious that during any two years' time, the variation of RPLs is considerable. Therefore, dividing the data sets into two years' segments and assuming that rotation rates within each of the segments are uniform is unlikely to be appropriate.

The altitude difference between the photosphere and the low corona is several thousand kilometers, about $\sim$1\% of the solar radius, and thus the RPLs of the three layers are likely identical. On the contrary, we found that they are generally differential.
It is possible that the three layers rotate with the same rate, and our result is affected by the different nature of the three indices. For instance, different lifetimes of objects responsible for the three indices, their different spatial scales, latitudinal distributions, and evolution processes, etc.
However, \citet{Li2019a} indeed found systematic patterns in the difference of RPL in different wavelengths of solar spectral irradiance from 1$\sim$2400 nm, which are generally formed in various depths of the solar atmosphere. So it is also possible that there is indeed intrinsic difference in solar rotation rates in various layers.
Further research is needed to investigate whether the differential rotation as a function of depth exists in the solar atmosphere.

The RPLs of all the three atmospheric layers have shown a decreasing trend during the last 40 years. The  variation of the RPLs at the recent four solar maxima confirms the decreasing trend. Though the variation of  RPLs-Mg at solar maxima is different, its overall trend agrees with the other two.
The lengths of the rotational period are declining, which means that the rotation of the solar atmosphere is accelerating.
This result agrees with previous studies on the solar photosphere and corona, e.g.,~\citet{Heristchi2009, Li2011b, Li2012, Xie2017a, Deng2020}; all of them have found decreasing trends in the photospheric or coronal rotation rates obtained with SN, F10.7 flux, or the modified coronal index.
The rotation of the chromosphere shows a larger accelerating rate than the photosphere and corona, and its rotation rates at solar maxima are different from those of the other two.

This overall secular trend is inferred to be linked with the level of solar magnetic activity. During the recent four SCs, the activity of the Sun also exhibits a declining trend as shown in Figure~\ref{fig-data}.
For decades, the relationship between solar rotation rate and solar activity level has been studied extensively.
For example, \citet{AiniKambry1990} studied the solar differential rotation with three solar cycles' sunspot drawings (1954--1986) produced at the National Astronomical Observatory of Japan, and they discovered that activity levels affect the rotation rate: the rotation rate at low activity cycle is larger than that at high activity cycle. In the same year, \citet{Hathaway1990a} inspected the sunspot rotation rate data (1921--1982) from Mount Wilson and concluded that, from cycle to cycle, the Sun rotates quicker when there are fewer sunspots and less sunspot area.
\citet{Obridko2001} studied that rotation of  large-scale global magnetic fields and its relation with local field activities during 1915--1996; they are found to be anti-correlated. The global fields' rotation rates decrease when the secular local field activity reaches maxima.
\citet{Brajsa2006} found an increasing trend in the solar rotation velocity with the sunspot group data during 1874--1981 and the residual method; they predict that the RPL should decline along with the decline of solar activity, based on the assumption that the Sun is at the end of a unusually high activity period.
\citet{Obridko2016} studied the relation between solar activity and solar rotation rate, and they noticed that a decrease in rotation rate occurs with an increase in solar activity.
Our result obtained from the chromospheric Mg II index confirmed such a relationship. This pattern should be taken into account in the construction of modern solar dynamo models.

By analyzing the significant RPLs with LSP and PDM methods, we found that all of them show periods that are inferred to be multiple harmonics of the Schwabe cycle.
\cite{Heristchi2009} found a 52.4 yr period mainly from the variation of the photospheric RPLs, which can not be determined here due to the limited length of the Mg II index data set.
The decreasing trend during the recent 40 years does not seem to be part of a 52.4 yr long-term cycle.

At last, we found that the variations of the three RPLs are modulated by the solar activity cycle.
The rotation rate variations of the photosphere, chromosphere, and corona lag the solar activity by $\sim$2.1 yr, $\sim$5.2 rotation periods, and  $\sim$11.3 rotation periods, respectively.
Besides, as mentioned previously, the variation of rotation rates at solar maxima is in accordance with the variation of the solar activity amplitudes.
\citet{Brajsa2006} also found that the dependence of the rotation rate on the phase of the solar cycle is noticeable.
 \citet{Obridko2016} uncovered that the RPL peaks 1--2 yr ahead of the SN maxima, which is at odds with our result (all of the three RPLs lag behind the SN). Strong magnetic fields seem to hinder the rotation. We think it is more likely that, because of the inertia of atmospheric plasma motion, the change of rotation rate should be led by the change of solar activity level.
In addition, contradictory results were yielded on the rotation rates at solar minima.
For example, \citet{Hathaway1990a}, \citet{Obridko2001}, and \citet{Brajsa2006} showed a faster rotation rate at solar minima, while \cite{Xie2017a} assert that the Sun rotates slower during solar minimum times. \citet{AiniKambry1990} even reached the conclusion that the equatorial rotation rate shows another systematic variation: it is high at the beginnings of an SC, and then decreases subsequently.
 As mentioned before, we reaffirm that it is unable to determine the rotation rates from solar indices during solar minima.
In all, more studies are necessary to tackle the internal-cycle and intra-cycle relationship between solar activity level and solar rotation rates.

\acknowledgments
We thank the anonymous referee for the helpful comments and suggestions.
The Bremen composite Mg II index is from \href{http://www.iup.uni-bremen.de/gome/gomemgii.html}{uni-bremen.de}.
The sunspot number is from \href{http://www.sidc.be/silso/datafiles}{WDC-SILSO}, Royal Observatory of Belgium, Brussels, and the F10.7 data set comes from the LASP Interactive Solar Irradiance Datacenter (\href{https://lasp.colorado.edu/lisird/data/penticton_radio_flux/}{LISIRD}).
This work is supported by the National Natural Science Foundation of China (No. 11903077, 11973085, 11633008, 40636031, 11673061, 11703085), the Basic Research Priorities Program of Yunnan (201901U070093), the Collaborating Research Program of CAS Key Laboratory of Solar Activity (KLSA201912, KLSA202012), the CAS "Light of West China" Program, and the Chinese Academy of Sciences.
This work made use of open-source software and packages including Numpy~\citep{Walt2011}, Scipy~\citep{Virtanen2020a}, Matplotlib~\citep{Hunter2007}, IPython~\citep{Perez2007},  Astropy, PyAstronomy, Seaborn, and Jupyter.

\bibliographystyle{aasjournal}
\bibliography{chrom_rotat}
\end{document}